\documentstyle[12pt,epsfig]{article}
\def\bge{\begin{equation}}
\def\ene{\end{equation}}
\def\bg{\begin{eqnarray}}
\def\en{\end{eqnarray}}
\def\nn{\nonumber}

\def\S0{{\Sigma^0}}

\def\k0bar{\bar{K}^0}

\begin{document}
\renewcommand{\thefootnote}{\fnsymbol{footnote}}
\begin{flushright}
ADP-99-19/T361
\end{flushright}
\begin{center}
{\LARGE Effect of nucleon structure variation on the longitudinal 
response function}
\end{center}
\vspace{0.5cm}
\begin{center}
\begin{large}
K.~Saito\footnote{Permanent address: Tohoku College of Pharmacy, 
Sendai 981-8558, Japan}\footnote{ksaito@nucl.phys.tohoku.ac.jp}, 
K.~Tsushima\footnote{ktsushim@physics.adelaide.edu.au} and 
A.W.~Thomas\footnote{athomas@physics.adelaide.edu.au} \\ 
\end{large} 
Special Research Center for the Subatomic Structure of Matter \\
and Department of Physics and Mathematical Physics \\
The University of Adelaide, SA 5005, Australia 
\end{center}
%
%
\begin{abstract}
Using the quark-meson coupling (QMC) model, we study 
the longitudinal response function for quasielastic electron 
scattering from nuclear matter.  In QMC the coupling constant between the 
scalar ($\sigma$) meson and the nucleon is expected to decrease with
increasing nuclear density, because of the self-consistent 
modification of the structure of the nucleon.  The reduction of the 
coupling constant then leads to a smaller contribution from
relativistic RPA than in the Walecka model.
However, since the electromagnetic form factors of the in-medium nucleon 
are modified at the same time, the longitudinal response function 
and the Coulomb sum are reduced by a total of about 20\% 
in comparison with the Hartree contribution.    
We find that the relativistic RPA and the nucleon 
structure variation 
both contribute about fifty-fifty to the reduction of 
the longitudinal response. 
\end{abstract}
\newpage

There is still considerable interest in the longitudinal response for 
quasielastic electron scattering.  Within the framework of nonrelativistic 
nuclear models and the impulse approximation,  
it is very difficult to reproduce the observed, quenched longitudinal 
response functions~\cite{exp}.  In the mid '80s, several groups calculated 
the longitudinal response function using Quantum Hadrodynamics 
(QHD)~\cite{qhd} (i.e., the Walecka model). They argued that 
the contribution of the relativistic random phase approximation (RRPA), 
which includes vacuum polarization, is very important in reducing the 
Coulomb sum rule~\cite{hor,several} below the sum of the 
squares of the nucleon charges in the nucleus. 
There have also been several other attempts to study the 
longitudinal response in nonrelativistic approaches~\cite{other}. 

On the other hand, the nucleon has internal structure, and it is nowadays 
expected that this structure should be modified in a nuclear environment 
\cite{change}. This is closely related to the issue of chiral
restoration in QCD. In QHD nuclear matter consists of 
{\em point-like\/} 
nucleons interacting through the exchange of point-like scalar 
($\sigma$) and vector ($\omega$) mesons.  It would clearly be very 
interesting to investigate the quenching of the longitudinal response 
function in a relativistic framework, including in addition, 
the structural changes of the nucleon in-medium.  

Recently, we have developed a relativistic quark model for nuclear matter, 
namely, the quark-meson coupling (QMC) model~\cite{qmc}, which could be 
viewed as an extension of QHD.  However, in QMC the mesons 
couple to confined quarks (not to point-like nucleons) and the 
nucleon is described by the MIT bag model.  This model yields an 
effective Lagrangian for a nuclear system~\cite{eff}, which has the same 
form as that in QHD with a 
{\em density dependent\/} coupling constant between the $\sigma$ and the 
nucleon (N) -- instead of a fixed value.  Indeed, from the point of view
of the energy of a nuclear system, the key difference 
between QHD and QMC lies in the $\sigma$-N coupling constant, $g_s$.  
Although this difference may seem subtle, it leads to many attractive 
results~\cite{qmc,eff}.  
We have already applied this model to various nuclear 
problems~\cite{apply}.  Here we use it to study  
the effect of nucleon structure variation in the 
longitudinal response function from nuclear matter.  

First, let us briefly review the calculation of the longitudinal 
response function for quasielastic electron scattering from (iso-symmetric) 
nuclear matter in QHD.  The starting point 
is the lowest order polarization insertion, $\Pi_{\mu \nu}$, for the 
$\omega$ meson.  This describes the coupling of a virtual vector meson 
or photon, of momentum $q$, to a particle-hole or nucleon-antinucleon 
excitation: 
\bge
\Pi_{\mu \nu}(q) = -ig_v^2 \int \frac{d^4k}{(2\pi)^4}
     \mbox{Tr}[G(k) \gamma_\mu G(k+q) \gamma_\nu], \label{piv}, 
\ene
where $G(k)$ is the self-consistent nucleon propagator (with momentum $k$) 
in relativistic Hartree approximation (RHA) given as 
\bg
G(k) &=& G_F(k) + G_D(k), \nn \\
   &=& (\gamma^\mu k^*_\mu + M^*) \left[ 
   \frac{1}{k^{*2}_\mu - M^{*2} + i\epsilon} + \frac{i\pi}{E^*_k} 
   \delta(k^*_0 - E_k^*) \theta(k_F - |{\vec k}|) \right].  
\label{propn}
\en
Here $k^{*\mu}=(k^0 - g_vV^0, {\vec k})$ ($V^0$ is the mean value of 
the $\omega$ field), $E_k^*=\sqrt{{\vec k}^2 + M^{*2}}$ ($M^*$ is 
the effective nucleon mass in matter) and $k_F$ is the Fermi momentum.  
Using the nucleon propagator we can separate the polarization insertion 
into two pieces: one is the density dependent part, $\Pi_{\mu \nu}^D$, 
which involves at least one power of $G_D$, and the other is the 
vacuum polarization insertion, $\Pi_{\mu \nu}^F$, which involves only 
$G_F$.  The former is finite, but the latter is divergent and 
must be renormalized.  We choose to renormalize such that 
$\Pi_{\mu \nu}^F(q)$ vanishes at $q_\mu^2 = m_\omega^2$ and $M^* = M$ 
(where $m_\omega$ and $M$ are respectively the free masses of the $\omega$ 
meson and the nucleon).  We then find~\cite{soprop} 
\bge
\Pi_{\mu \nu}^F(q) = \xi_{\mu \nu} \Pi^F(q), 
\label{vcond}
\ene
with $\xi_{\mu \nu} = - g_{\mu \nu} + (q_\mu q_\nu /q_\mu^2)$ and 
\bg
\Pi^F(q) &=& \frac{g_v^2}{6\pi^2} q_\mu^2 \biggl[ 
2 \ln\frac{M^*}{M} - 4\left( \frac{M^{*2}}{q_\mu^2} - 
\frac{M^2}{m_\omega^2} \right)  \\ \nn
&+& \left( 1 + 2\frac{M^{*2}}{q_\mu^2} \right) 
f(x_q) - 
\left( 1 + 2\frac{M^2}{m_\omega^2} \right) 
f(z_v) \biggr],  \label{v-loop}
\en
where $x_q=1-\frac{4M^{*2}}{q_\mu^2}$, $z_v=1-\frac{4M^2}{m_\omega^2}$ 
and 
\bge
f(y) = \left\{ 
\begin{array}{rl}
\sqrt{y} \ln\frac{\sqrt{y}+1}{\sqrt{y}-1}, & \ \ \ 
\mbox{ for $1 \leq y < +\infty$} \\
\sqrt{y} \ln\frac{1+\sqrt{y}}{1-\sqrt{y}} -i \pi \sqrt{y}, & \ \ \
\mbox{ for $0 < y < 1$} \\ 
2\sqrt{-y} \tan^{-1}\frac{1}{\sqrt{-y}}. & \ \ \
\mbox{ for $y \leq 0$} \end{array} \right.  \label{fff}
\ene
We assume that the isospin degeneracy of the vacuum is 2.   For 
$\Pi_{\mu \nu}^D$, the explicit, analytical expressions can be 
found in Ref.~\cite{lim} (also see Ref.~\cite{soprop}).  

In the Hartree approximation, where only the lowest one nucleon ring is 
considered, the longitudinal response function, 
$S_L^H$, measured in electron scattering is simply given by 
\bge
S_L^H(q) = - \left( \frac{Z G_{pE}^2(q) |{\vec q}|^2}{g_v^2 
\pi \rho_B q_\mu^2} \right) {\Im}m \Pi_L(q). 
\label{sl}
\ene
Here $Z$ is the nuclear charge, $\rho_B$ the nuclear density, 
$\Pi_L (= \Pi_{33} - \Pi_{00})$ the longitudinal component of the 
polarization insertion (we choose the direction of ${\vec q}$ as 
the $z$-axis) and $G_{pE}$ is the proton electric form factor,   
which is usually parametrized by a dipole form in free space: 
\bge
G_{pE}(Q^2) = \frac{1}{(1 + Q^2/0.71)^2},   
\label{gpe}
\ene
with the space-like momentum transfer, $Q^2 = - q_\mu^2$, in units of 
GeV$^2$.  For this initial investigation we omit  
a small (and rather complicated) contribution from the 
anomalous moments~\cite{hor}, in order to concentrate on the role  
of the variation of the structure of the nucleon.  
Since the vacuum polarization is real in the space-like region there is no 
modification of the Hartree response from this term.  

The RRPA for the longitudinal component of the polarization insertion, 
$\Pi_L^{RPA}$, involves the sum of the ring diagrams to all orders.  This 
summation has been discussed by many 
authors~\cite{hor,several,soprop,lim,ring}. 
It involves $\sigma$-$\omega$ mixing in the nuclear medium, and 
is given by 
\bge
\Pi_L^{RPA}(q) = [(1 - \Delta_0 \Pi_s) \Pi_L + \Delta_0 \Pi_m^2] 
/ {\epsilon_L} , 
\label{rrpa}
\ene
where $\epsilon_L$ is the longitudinal dielectric function 
\bge
\epsilon_L = (1 - d_0 \Pi_L)(1 - \Delta_0 \Pi_s) - 
(q_\mu^2/q^2) \Delta_0 d_0 \Pi_m^2, \label{long}
\ene
with $q=|{\vec q}|$, and the free meson propagators for the $\sigma$ and 
$\omega$ mesons are respectively 
\bge
\Delta_0(q) = \frac{1}{q_\mu^2 - m_\sigma^2 + i\epsilon}  \ \ \ 
\mbox{and} \ \ \ 
d_0(q) = \frac{1}{q_\mu^2 - m_\omega^2 + i\epsilon},  
\label{freeo}
\ene
where $m_\sigma$ is the $\sigma$ meson mass.  
Here $\Pi_s$ and $\Pi_m$ are respectively the scalar and the time 
component of the mixed polarization insertions: 
\bg
\Pi_s(q) &=& -ig_s^2 \int \frac{d^4k}{(2\pi)^4}\mbox{Tr}[G(k)G(k+q)],  
\label{pis} \\
\Pi_m(q) &=& ig_sg_v \int \frac{d^4k}{(2\pi)^4}
     \mbox{Tr}[G(k) \gamma^0 G(k+q)]. 
\label{pim}
\en
The scalar polarization insertion can be again separated into two pieces.  
The density dependent part is finite and the explicit expression 
can be found in Ref.~\cite{lim}. Because it does not 
involve $G_D$, the vacuum component, $\Pi_s^F$, is, of course, divergent 
and once again we need to renormalize it.  First, 
we introduce the usual counter terms to the Lagrangian, 
which includes terms quadratic, cubic and quartic in the 
$\sigma$ field, as well as wavefunction renormalization~\cite{qhd}.  
To get the ``physical'' properties of the $\sigma$ meson in free space, 
we impose the following condition~\cite{soprop}: 
\bge
\Pi_s^F(q_\mu^2, M^*=M) = 
\frac{\partial}{\partial q_\mu^2} \Pi_s^F(q_\mu^2, M^*=M) = 0 
\ \ \ \mbox{ at } q_\mu^2 = m_\sigma^2.  \label{scond}
\ene
Then, we find 
\bg
\Pi_s^F(q) &=& \frac{3g_s^2}{2\pi^2} \biggl[ 
\frac{1}{6} (m_\sigma^2 - q_\mu^2)  \\ \nn
&-& \left( M^{*2} - \frac{q_\mu^2}{6} \right) \left( 2 
\ln\frac{M^*}{M} 
+ f(x_q) -  f(z_s) \right)  \\  \nn
&+& \frac{q_\mu^2}{3} \left( \frac{M^{*2}}{q_\mu^2} 
(f(x_q)-2) - \frac{M^2}{m_\sigma^2} 
(f(z_s)-2) \right) \\ \nn
&-& (M^{*2} - M^2) ( f(z_s)-2) 
+ 2M(M^*-M) + 3(M^*-M)^2 \biggr],  \label{n-loop}
\en
where $z_s=1-\frac{4M^2}{m_\sigma^2}$.  
For the mixed polarization insertion there is no vacuum polarization and 
it vanishes at zero density.  (The explicit form can be also found in 
Ref.~\cite{lim}.) 

As QHD involves only isoscalar mesons, the 
isovector RRPA response is the same as the Hartree one, eq.(\ref{sl}). 
This implies that the vacuum polarization only affects the isoscalar 
response.  It remains to study the effect of isovector mesons.  
(In the isovector part the rho meson coupling (without vacuum 
polarization) was studied in Ref.~\cite{rho}.  It reduces $S_L$ slightly.) 
Since the longitudinal response is half isoscalar and half isovector, 
the longitudinal response function in RRPA is given by~\cite{hor}
\bge
S_L^{RPA}(q) = - \left( \frac{Z G_{pE}^2(q) |{\vec q}|^2}{g_v^2 
\pi \rho_B q_\mu^2} \right)  
{\Im}m \left[ \frac{\Pi_L^{RPA}(q) + \Pi_L(q)}{2} \right]. 
\label{slrpa}
\ene
Several authors~\cite{hor,several} have calculated the longitudinal 
response function using this RRPA polarization, and 
reported that it is very important in reproducing the observed 
experimental data~\cite{exp}.  

Now we are in a position to discuss the effect of changes in the 
internal structure of 
the nucleon in-medium. In order to do so, we consider the following 
modifications to the QHD approach:   

\noindent(1) meson-nucleon vertex form factor 
 
In QHD the interactions between the mesons and nucleon are {\em point-like}.  
However, since both the mesons and nucleon are composite they have finite 
sizes.  In the region of space-like momentum transfer the finite-size 
effect reduces the meson-N coupling. As the simplest example, one could
take a monopole form factor~\cite{bonn} at each vertex:
\bge
F_N(Q^2) = \frac{1}{1 + Q^2/\Lambda_N^2} ,
\label{formn}
\ene
with a cut off parameter $\Lambda_N = 1.5$ GeV.  In principle, 
one could self-consistently calculate the form factor within QMC.
However, as such changes are not expected to make a big difference, we 
use eq.(\ref{formn}) in the following calculation.  

\noindent(2) modification of the proton electric form factor 

Recently we have studied the electromagnetic form 
factors of the nucleon, not only in free space~\cite{emff} but also in a 
nuclear medium, using the QMC model~\cite{emffm}  
(see also Ref.~\cite{berg}). 
Because the confined quark feels an attractive force due to the $\sigma$,  
the quark wave function is modified in a nuclear medium.  
The ratio of the electric form factor of the proton in medium to that 
in free space, $G_{pE}(\rho_B,Q^2)/G_{pE}(Q^2)$, is shown in Fig.3 in 
Ref.~\cite{emffm}.  From the figure we can see that the ratio 
decreases very linearly as a function of $Q^2$, and that it is accurately  
parametrized at $\rho_B = \rho_0$ (= 0.15 fm$^{-3}$, the normal nuclear 
matter density) as 
\bge
R_{pE}(\rho_0,Q^2) \equiv \frac{G_{pE}(\rho_0,Q^2)}{G_{pE}(Q^2)} \simeq  
1 - 0.26 \times Q^2. \label{gapp}
\ene
This implies that the 
(electric) rms radius of the proton at $\rho_0$ swells by about 5.5 
\% (for more details, see Ref.~\cite{emffm}).  
Since the bag model reproduces the form factor measured in free space 
very well~\cite{emff} and the latter is
well described by eq.(\ref{gpe}), 
the in-medium proton form factor can be represented as 
$G_{pE}(Q^2) \times R_{pE}(\rho_B,Q^2)$\footnote{
The nucleon-antinucleon excitation in $\Pi_{\mu \nu}$ contributes to the
photon-nucleon vertex as a RRPA correction, which may 
in principle lead to double
counting for the form factor.  However, we ignore this correction 
because it is very small~\cite{several}.
}.  

\noindent(3) density dependence of the coupling constants 

In QMC the confined quark in the nucleon couples to the $\sigma$ field
which gives rise to an attractive force.
As a result the quark becomes more 
relativistic in a nuclear medium than in free space.  This implies that 
the small component of the quark wave function, $\psi_q$, is enhanced in 
medium~\cite{qmc,eff}.  
The coupling between the $\sigma$ and nucleon is therefore expected to be 
reduced at finite density because it is given in terms of the quark scalar 
charge, $\int_{Bag} dV\/{\bar \psi}_q \psi_q$~\cite{eff}.  
On the other hand, the coupling between the vector meson and nucleon 
remains constant, because it is related to the baryon number, which is
conserved.  

To study the longitudinal response of nuclear matter, we first have 
to solve the nuclear ground state within RHA.  
In QHD the total energy density for nuclear matter is written 
as~\cite{soprop} 
\bge
{\cal E} = {\cal E}_0 + \frac{1}{2\pi^2}M^2(M-M^*)^2 \left[ 
  \frac{m_\sigma^2}{4M^2} + \frac{3}{2} f(z_s) 
-3 \right], \label{energy} 
\ene
where ${\cal E}_0$ has the usual form (in RHA), given in Ref.~\cite{qhd}.  
Note that in Ref.~\cite{qhd} the renormalization condition on the nucleon 
loops is imposed at $q_\mu^2$=0. The second term on the 
r.h.s. of eq.(\ref{energy})~\cite{soprop} occurs because we chose
the renormalization condition for the $\sigma$ at $q_\mu^2=m_\sigma^2$
(see eq.(\ref{scond})). As measureable quantities cannot depend on this
choice, our 
model gives the same physical quantities as those of Ref.~\cite{qhd}.  

To take into account the modifications (1) and (3), we replace the 
$\sigma$- and $\omega$-N coupling constants in eq.(\ref{energy}) by 
\bg
g_s &\to& g_s(\rho_B) \times F_N(Q^2),  \label{replace1}  \\
g_v &\to& g_v \times F_N(Q^2),  \label{replace2}
\en
where the density dependence of $g_s(\rho_B)$ is given by solving the 
nuclear matter problem self-consistently, using the 
MIT bag for the nucleon model (see Ref.~\cite{eff}).  
As in QHD, we have two adjustable parameters in the present calculation: 
$g_s(0)$ (the $\sigma$-N coupling constant at $\rho_B=0$) and $g_v$.    

Requiring 
the usual saturation condition for nuclear matter,  
namely ${\cal E}/\rho_B - M = - 15.7$ MeV at 
$\rho_0$, we determine the coupling constants $g_s^2(0)$ and $g_v^2$  
($g_s^2(0)=61.85$ and $g_v^2=62.61$).  
In the calculation we fix the quark mass to be 5 MeV, $m_\sigma=550$ MeV 
and $m_\omega=783$ MeV, while the bag parameters are chosen so as 
to reproduce the free nucleon mass ($M=939$ MeV) with the 
bag radius $R_0=0.8$ fm (i.e., $B^{1/4}=170.0$ MeV and 
$z=3.295$~\cite{qmc,eff}). 
This yields the effective nucleon mass $M^*/M=0.81$ at $\rho_0$ 
and the incompressibility $K=281$ MeV.  
(We do not consider the 
possibility of medium modification of the meson properties~\cite{eff}   
in the present work.)  

\begin{figure}[htb]
\begin{center}
\epsfig{file=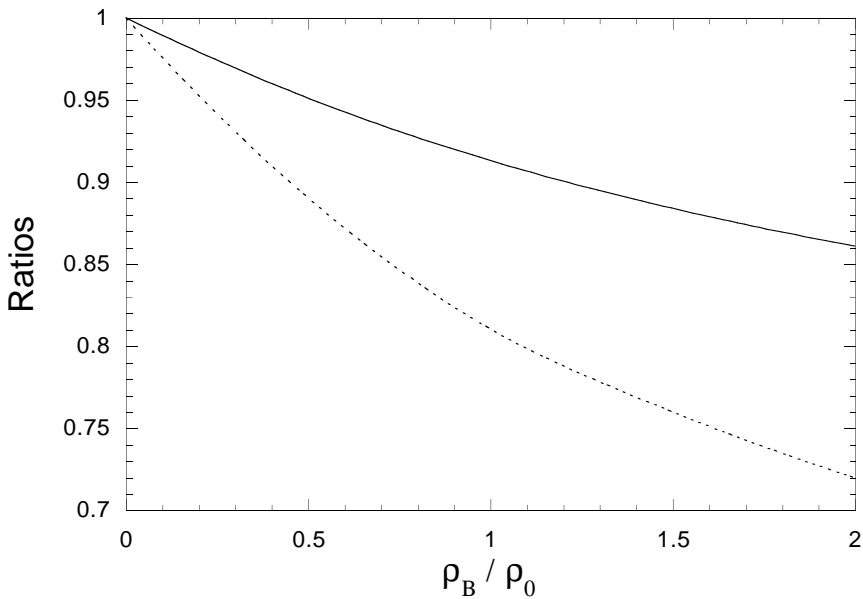,height=8cm}
\caption{Density dependences of $g_s(\rho_B)/g_s(0)$ and $M^*/M$.  
The solid curve is for the ratio of the coupling constants, while 
the dotted curve is for the ratio of the nucleon masses. 
}
\label{f:ratios}
\end{center}
\end{figure}
\begin{figure}[htb]
\begin{center}
\epsfig{file=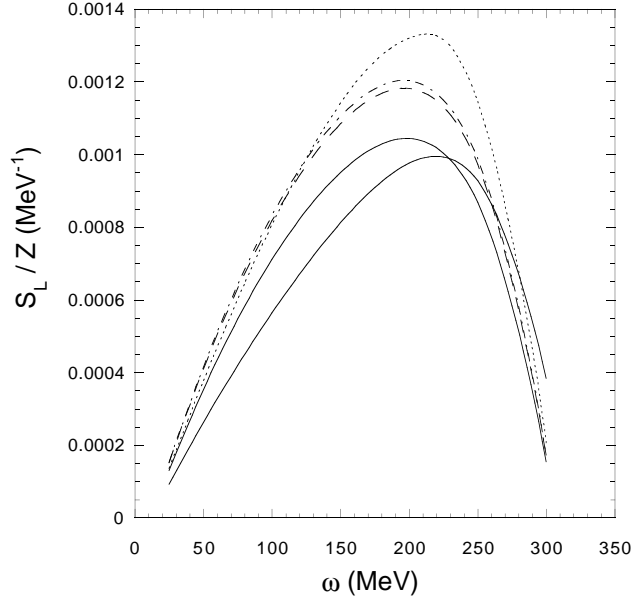,height=8cm}
\caption{Longitudinal response functions in QMC with $m_q$ = 
5 MeV.  We fix $q$ = 550 MeV and $\rho_B = \rho_0$.   
The dotted curve is the result of the Hartree approximation 
(see eq.(\protect\ref{sl})), where the effective nucleon mass is given by 
QMC and the proton electric form factor is the same as in free space.  
The dashed curve is the result of the full 
RRPA, without the modifications (1) and (2) (i.e. $F_N=1$ and 
$R_{pE}=1$).  
The dot-dashed curve shows the 
result of the full RRPA with the meson-N form factor but $R_{pE}=1$.  
The upper (lower) solid curve shows the result of the full RRPA for 
$m_q = 5~(300)$ MeV, including all modifications.  
}
\label{f:resp}
\end{center}
\end{figure}
\begin{figure}[htb]
\begin{center}
\epsfig{file=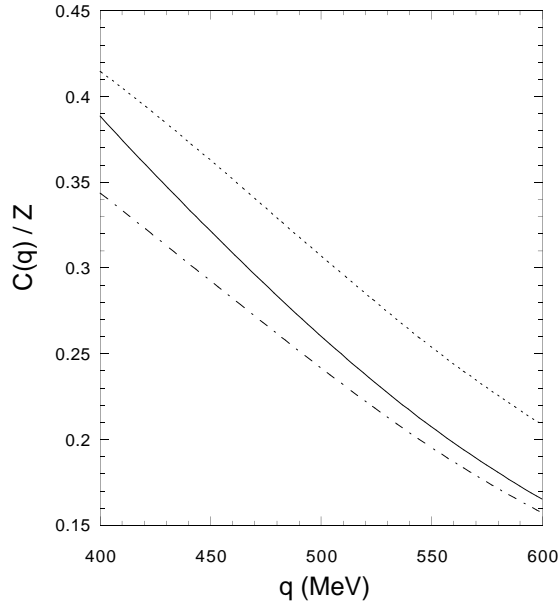,height=8cm}
\caption{Coulomb sum, $C(q)/Z$, at $\rho_0$ 
in the Hartree approximation 
with $R_{pE}=1$ (the dotted curve) or the full RRPA with 
all modifications (the solid and dot-dashed curves are 
for $m_q$ = 5 and 300 MeV, respectively).  
}
\label{f:csr}
\end{center}
\end{figure}
Now we present our main results.  First, in Fig.~\ref{f:ratios}, we show 
the density dependence of
the coupling constant.   At $\rho_0$, $g_s$ decreases by about 9\%.
The effective nucleon mass is also shown in the figure. 

Next, we show the longitudinal response function in QMC.  
Using the density dependent coupling constant, the meson-N form factors  
(see eqs.(\ref{replace1}) and (\ref{replace2})) and the in-medium proton 
electric form factor, we can calculate the 
longitudinal response of nuclear matter.  The result is shown in 
Fig.~\ref{f:resp}.  Because of the density dependent coupling, 
$g_s(\rho_B)$, the reduction of the response function due to the full 
RRPA (the dashed curve in the figure) from the Hartree result (the 
dotted curve) becomes much smaller than that in QHD.   
On the other hand, 
the modification of the proton electric form factor is very 
significant, yielding a much bigger reduction in the response 
(see the upper solid curve).  
We can see that the effect of the meson-N form factor enhances the 
longitudinal response (see the dot-dashed curve), but it is not large. 

It is also interesting to see the quark mass dependence of the 
longitudinal response.  As an example, we consider the case of $m_q = 
300$ MeV, which is a typical constituent quark mass.  For $m_q$ = 300 MeV 
and $R_0$ = 0.8 fm, the coupling constants required to fit the 
saturation properties of nuclear matter are: $g_s^2(0)$ = 68.69 and 
$g_v^2$ = 84.24, and the effective nucleon mass at $\rho_0$ and the 
incompressibility become 723 and 345 MeV, respectively.  Using these 
parameters we show the result for the longitudinal response (the lower 
solid curve) in Fig.~\ref{f:resp}.   
In comparison with the case $m_q$ = 5 MeV, 
it is a little smaller and the peak position is shifted to the higher 
energy transfer side.  This may be due to the smaller effective nucleon 
mass in the case $m_q$ = 300 MeV than when $m_q$ = 5 MeV.  

The integrated strength of the longitudinal response  
(or the Coulomb sum), $C(q)$, 
\bge
C(q) = \int^q_0 dq_0 \/ S_L(q, q_0),    \label{coulomb}
\ene
is shown in  Fig.~\ref{f:csr} as a function of three-momentum transfer, 
$q$.  For high $q$, the strength is about 20\% lower in the full 
calculation than for the Hartree response.  For low $q$, the full 
calculation with the constituent quark mass remains much lower 
than the Hartree result, while in case of the light quark mass it 
gradually approaches the Hartree one.  This difference is caused by that 
the effective nucleon mass for $m_q$ = 5 MeV being larger in matter than that 
for $m_q$ = 300 MeV.  
The 20\% reduction found here is a little smaller than the 
value of approximately 30\% found in QHD~\cite{hor,several}.  

We would like to emphasize that these calculations are for nuclear matter  
and cannot be directly compared with the experimental data. Furthermore, 
there still remain discrepancies and 
uncertainties in the present experimental results~\cite{jordan,new}.  

{}Finally, we comment on the transverse response from nuclear 
matter.  In Ref.~\cite{emffm} we can see that in QMC 
the modification of the nucleon magnetic form factor 
in-medium is very small: the calculated decrease in the proton (neutron) 
magnetic form factor is about 1.5\% (0.9\%) at $\rho_0$.  Therefore, 
one would expect the total change in the transverse response caused by  
RRPA correlations and the effect of the variation of the structure of
the nucleon to be much smaller than in the longitudinal response.
This is certainly what one needs in order to fit 
the experimental data~\cite{exp}.  

In summary, we have calculated the longitudinal response of nuclear 
matter using the QMC model.  The reduction of the 
$\sigma$-N coupling constant with density decreases the 
contribution of the RRPA, while the modification of the proton electric 
form factor in medium reduces the longitudinal response 
considerably.  The longitudinal response, or the Coulomb sum, is  
reduced by about 20\% in total, with RRPA correlations and the variation 
of the in-medium nucleon structure contributing about fifty-fifty.  
In the near future we hope to extend this work to calculate the 
longitudinal and transverse 
response functions for finite nuclei, using local density approximation, 
in order to compare our results directly with new experimental 
results~\cite{new}.  

\vspace{1cm}
K.S. would like to acknowledge the warm hospitality at the CSSM, where 
this work was partly carried out.  K.S. would especially like to  
thank M. Ericson, P.A.M. Guichon, W. Bentz and 
J. Morgenstern for valuable discussions on the longitudinal response 
function.  
This work was supported by the Australian Research Council and 
the Japan Society for the Promotion of Science.  

\clearpage
%
%

%
\end{document}